\documentstyle[amsmath,amssymb,graphicx]{article}

\def\be{\begin{eqnarray}}
\def\ee{\end{eqnarray}}
\def\nn{\nonumber}

\def\l[{\phantom.[}

\hoffset= - 1.1in         

\begin{document}



\noindent
{\it to Branko Dragovich} \hfill IITP/TH-12/15 \\
{\it on his 70'th birthday}

\bigskip

\centerline{\Large{Are there $p$-adic knot invariants?
}}

\bigskip

\centerline{{\bf  A.Morozov }}

\bigskip

{\footnotesize
\centerline{{\it
ITEP, Moscow 117218, Russia}}

\centerline{{\it
Institute for Information Transmission Problems, Moscow 127994, Russia
}}

\centerline{{\it
National Research Nuclear University MEPhI, Moscow 115409, Russia
}}

\bigskip

\centerline{ABSTRACT}

\bigskip

{\footnotesize
We suggest to use the Hall-Littlewood version \cite{HL} of Rosso-Jones formula
to define the germs of $p$-adic HOMFLY-PT polynomials for torus knots $[m,n]$,
which possess at least the $[m,n] \longleftrightarrow [n,m]$ topological invariance.
This calls for generalizations to other knot families
and is a challenge for several branches of modern theory.
}

\bigskip

\bigskip

$P$-adic strings got a lot of attention in  late 80's \cite{pstr},
after adelic property of $\Gamma$-functions
was applied to the 4-point tree Veneziano amplitude \cite{adelVen}.
However, the failure of naive adelization of loop amplitudes \cite{LeMo}
along with other similar problems moved the entire subject away from the
proscenium of theoretical physics (attention to it is now partly shifted
to applications in biology and computer science)
-- what is unjust, because it was clearly demonstrated that quantum field theory
is very well suited to $p$-adization.
In fact, recent advances in conformal and knot theories open a new window
of possibilities for $p$-adic physics, and this opportunity should not be missed.
There are now numerous points, where it should emerge in quite a non-trivial way,
and this short letter is an attempt to attract attention to the subject.

\bigskip

The story can begin from different starting points, for example, from a simple identity
\be
\frac{N(N+1)}{2} - \frac{N(N-1)}{2}=
\frac{N(N+1)(N+2)}{6} - \frac{N(N-1)(N+1)}{3}+ \frac{N(N-1)(N-2)}{6}
\label{23cl}
\ee
which has a straightforward quantum deformation:
\be
t^{3N}\cdot\left(t^{-3}\cdot \frac{[N][N+1]}{[2]}  -  t^{3}\cdot\frac{[N][N-1]}{[2]}\right)=
t^{4N}\cdot\left(
t^{-4}\cdot \frac{[N][N+1][N+2]}{[2][3]} -  \frac{[N][N-1][N+1]}{[3]}
 +  t^{4}\cdot\frac{[N][N-1][N-2]}{[2][3]}\right)
\label{23id}
\ee
where $[M] = \frac{t^M-t^{-M} }{t-t^{-1}}$ is the "quantum number".

An important interpretation of  (\ref{23id}) is the Reidemeister invariance
of the HOMFLY-PT polynomial \cite{knotpols}
$H_{\Box}^{3_1}(t|A=t^N)$ for the simplest possible knot --
trefoil $3_1$, represented as either as $[2,3]$ or as $[3,2]$ torus knot.
Accordingly (\ref{23id}) has infinitely many generalizations of different level
of complexity -- starting from equivalence of fundamental HOMFLY for $[m,n]$ and $[n,m]$
torus knots and ending by equivalence between colored HOMFLY for various braid representations
of arbitrarily complicated satellite knots.
For modern approaches to the problem see \cite{RTmod} and references at \cite{knotebook}.

A natural question is if (\ref{23cl}) and its generalizations can possess not only
quantum, but also $p$-adic deformations.
It would be natural to expect such possibility, because HOMFLY polynomials are
Wilson loop averages in Chern-Simons theory \cite{CS,Wit,MSm}, which are closely related
\cite{Wit,inds} to conformal blocks \cite{CFT} in WZNW model \cite{WZNW} --
which in turn possesses a free field representation \cite{GMMOS},
while free fields have a natural $p$-adic counterpart in terms of a quantum field theory
on Bruhat-Tits trees \cite{FZ}.
Surprisingly or not, this question did not yet attract sufficient attention --
and not much is known about $p$-adic counterparts of conformal blocks,
solutions to BPZ and KZ equations, their modular transformations, and knot polynomials.
This letter is an attempt to attract attention to this circle of problems.
We shall follow here only one line of reasoning, which goes directly from identities
like (\ref{23id}) and seems most related to the
hard questions in modern knot theory.
However, by no means the more direct approaches, mentioned
earlier in this
paragraph, are less important and interesting.

The most naive thing to do would be to rewrite the representation dimensions,
actually appearing in (\ref{23cl}) and (\ref{23id}) as ratios of Gamma-functions
and $p$-deform them by the usual rule
\be
\Gamma(x) \ \longrightarrow \ \Gamma_p(x) =\frac{1-p^{x-1}}{1-p^{-x}}
\ee
This is, however, too naive -- and does not work
\ (in fact, more interesting from this perspective should be expressions for colored HOMFLY
{\it a la} \cite{IMMMfe}, which involve Pochhammer symbols and satisfy remarkable
recursions \cite{Gar}, closely related to the AMM/EO topological recursion \cite{AMMEO}).

\bigskip

A less naive approach is to interpret (\ref{23id}) as an identity between characters
$\chi_R\{p_k\}$
at the topological locus
\be
p_k = p_k^*\equiv \frac{A^k-A^{-k}}{t^k-t^{-k}} = \frac{[Nk]}{[k]},
\ \ \ \ \ \ \ A=t^N
\label{tolo}
\ee
-- as suggested in \cite{DMMSS}:
\be
A^3\cdot\Big(t^{-3}\cdot \chi_{[2]}^* \ -\ t^{3}\cdot\chi_{[11]}^*\Big) \ = \
A^4\cdot\Big(t^{-4} \cdot \chi^*_{[3]} \ - \ \chi^*_{[21]} \ + \ t^{4}\cdot\chi^*_{[111]}\Big)
\label{RJ2332}
\ee
The two sides of this equality are just the Rosso-Jones formulas \cite{RJ} for the torus HOMFLY
$[2,3]$ and $[3,2]$, if the role of characters is played by the ordinary Schur functions.

What is less trivial, (\ref{RJ2332}) can be lifted to an identity, involving MacDonald polynomials
\cite{MD}, which depend not only on the time variables $\{p_k\}$, but also  on $t$
and on additional parameter $q$.
According to \cite{DMMSS},
\be
\left(\frac{qA}{t}\right)^3\!\cdot \Big(
q^{-3} \cdot c_{[2]}M_{[2]}^* \ -\ t^{3}\cdot c_{[11]}M_{[11]}^* \Big)\ = \
\left(\frac{qA}{t}\right)^4\!\cdot \Big(
q^{-4} \cdot c_{[3]}M^*_{[3]} \ - \   c_{[21]}M^*_{[21]} \ +
\ t^{4}\cdot c_{[111]}M^*_{[111]}\Big)
\label{23MD}
\ee
This identity expresses Reidemeister invariance of {\it super}polynomials \cite{suppo} --
in the same case of $[2,3]$ and $[3,2]$ braid realizations of the torus trefoil.
MacDonald polynomials here are
\be
M_{[2]} = \frac{(1+t^2)(1-q^2)}{2(1-q^2t^2)}\cdot p_2 + \frac{(1-t^2)(1+q^2)}{2(1-q^2t^2)}\cdot p_1^2,
\ \ \ \ \ \ \ \ \ \ \ \
M_{[11]}= -\frac{1}{2}\cdot p_2 + \frac{1}{2}\cdot p_1^2, \nn \\ \nn \\
M_{[3]} = \frac{(1-q^2)(1-q^4)(1-t^6)}{3(1-t^2)(1-q^2t^2)(1-q^4t^2)}\cdot p_3 +
\frac{(1-t^4)(1-q^6)}{2(1-q^2t^2)(1-q^4t^2)}\cdot p_2p_1 +
\frac{(1+q^2)(1-q^6)(1-t^2)^2}{6(1-q^2)(1-q^2t^2)(1-q^4t^2)}\cdot p_1^3, \nn\\
M_{[21]} = -\frac{(1-q^2)(1-t^6)}{3(1-t^2)(1-q^2t^4)}\cdot p_3 +
\frac{(1+t^2)(t^2-q^2)}{2(1-q^2t^4)}\cdot p_2p_1 +
\frac{(1-t^2)(2+q^2+t^2+2q^2t^2)}{6(1-q^2t^4)}\cdot p_1^3,
\nn \\
M_{[111]}= \frac{1}{3}\cdot p_3 - \frac{1}{2}\cdot p_2p_1 + \frac{1}{6}\cdot p_1^3
\ee
with $\{x\} = x-x^{-1}$,
and an extra complication in (\ref{23MD}) is appearance of  non-trivial coefficients
\be
c_{[2]}=1, \ \ \ \ c_{[11]}=\frac{\{t^2\}}{\{qt\}},
\ \ \ \ \ \ \ \ \ \ \ \ \ \ \ \
c_{[3]}=1, \ \ \ \
c_{[21]}= \frac{\{t\}\cdot(q^{-2}+t^2q^{-2}+t^2)}{\{q^2t\}}, \ \ \ \
c_{[111]}= \frac{\{t^2\}\{t^3\}}{\{qt\}\{qt^2\}}
\ee

\bigskip

Now comes the crucial point -- as discovered in \cite{HL},
the overloaded expressions for superpolynomials at the two sides of (\ref{23MD})
simplify drastically, if rewritten through Hall-Littlewood polynomials
$L_R\{t|p_k\} = M_R\{q=0,t|p_k\}$:
\be
\boxed{
P^{[2,3]}_{_\Box} =
(A/t)^3\cdot \Big(
L_{[2]}^*(t) + q^2(1-t^4)\cdot L_{[11]}^*(t)\Big)\ = \
(A/t)^4\cdot\Big(
L_{[3]}^*(t) + q^2(1-t^2)\cdot L_{[21]}^*(t)\Big)
= P^{[3,2]}_{_\Box}
}
\label{23HL}
\ee
where
\be
L_{[2]} = \frac{1+t^2}{2}\cdot p_2 + \frac{1-t^2}{2}\cdot p_1^2, \ \ \ \ \ \ \ \ \ \
L_{[11]} = -\frac{1}{2}\cdot p_2 + \frac{1}{2}\cdot p_1^2\nn \\
\nn \\
L_{[3]} = \frac{1+t^2+t^4}{3}\cdot p_3 + \frac{1-t^4}{2}\cdot p_2p_1 + \frac{(1-t^2)^2}{6}\cdot p_1^3,
\nn\\
L_{[21]} =  -\frac{1+t^2+t^4}{3}\cdot p_3 + \frac{t^2(1+t^2)}{2}\cdot p_2p_1 + \frac{(1-t^2)(2+t^2)}{6}\cdot p_1^3,
\nn \\
L_{[111]}= \frac{1}{3}\cdot p_3 - \frac{1}{2}\cdot p_2p_1 + \frac{1}{6}\cdot p_1^3
\label{Lpols}
\ee
and
\be
L_{[2]}^*= \frac{t}{A}\cdot\frac{\{A\}}{\{t\}}, \ \ \ \ \ \ \ \ \
L_{[3]}^* = \frac{t^2}{A^2}\cdot\frac{\{A\}}{\{t\}}, \ \ \ \ \ \ \ \ \
L_{[11]}^* = \frac{\{A\}\{A/t\}}{\{t\}\{t^2\}}, \ \ \ \ \ \ \ \ \
L_{[21]}^* = \frac{t^2}{A}\cdot\frac{\{A\}\{A/t\}}{\{t\}^2}
\label{Ldims}
\ee

\bigskip

\noindent
For general torus knots the sum in the Hall-Littlewood representation of Rosso-Jones
formula goes not over all Young diagrams $Q$ where the number of boxes $|Q|$
is equal to the number of strands $m$, but the number of lines in $Q$ is also restricted
by the remainder $r$ in $n=mk+r$: $\ l(Q)\leq r$.
To emphasize the different role of $q$ and $t$ in (\ref{23MD}) we mention that the limit
$t=0$ (with $p_k$ fixed) is very different from (\ref{23HL}) -- and, perhaps,
even more interesting, see \cite{HL} for  further details.

\bigskip

What is important for our narrow purpose in this letter,
the deformation of (\ref{23cl}) into (\ref{23HL})
is not only simple and beautiful -- it also possesses a $p$-adic interpretation.
The point is that the Hall-Littlewood limit ($q=0$) of MacDonald polynomials is the one,
which describes zonal spherical functions (radial solutions to Laplace-Beltrami equations)
on $p$-adic homogeneous spaces \cite{phs}, it is enough to put $t=1/p$.
In \cite{FZ} it was therefore suggested to consider $M_R\{q=0,t=p^{-1}|p_k\}$
as the proper $p$-adic substitutes of the Schur functions.
Following this logic we can call the quantity at the two sides of (\ref{23HL})
evaluated at $t=1/p$ the $p$-adic HOMFLY polynomial -- with (\ref{23HL}) guaranteeing
its topological invariance (at least partial).
Since $L_{[s]}^*(t)\sim\frac{\{A\}}{\{t\}}$ in pure symmetric representations $[s]$
are practically independent of $s$ (and all provide one and the same answer for the unknot),
in order to get something non-trivial one should better pick up the first non-trivial germ
$\left.dP/dq^2\right|_{q=0}$, or, perhaps, consider the entire collection of coefficients
in front of different powers of $q$ (which altogether constitute the superpolynomial --
but are naturally graded by powers of $q$).

\bigskip

In the same way one can define $p$-adic HOMFLY for many other torus knots,
for example
\be
(t/A)^n\cdot P^{[2,n]}_{_\Box}\ = \
L_{[2]}(t)\ +\ q^2(1-t^4)\cdot\frac{1-(qt)^{n-1}}{1-q^2t^2}\cdot L_{[11]}(t)
\ee
and
\be
(t/A)^{8}\cdot P^{[3,4]}_{_\Box}\ = \ L_{[3]} (t) \ +\ q^2(1-t^2)(1+q^2+q^2t^2)\cdot L_{[21]} (t)
\ + \  q^6(1-t^4)(1-t^6)(1+q^2t^2)\cdot L_{[111]} (t) \nn \\ \nn\\
(t/A)^{10}\cdot P^{[3,5]}_{_\Box}\ = \ L_{[3]} (t) \ +\ q^2(1-t^2)(1+q^2)(1+q^2t^2)\cdot L_{[21]} (t)
\ + \  q^6(1-t^4)(1-t^6) \cdot L_{[111]} (t) \nn
\ee
\be
(t/A)^{14}\cdot P^{[3,7]}_{_\Box}\ = \ L_{[3]} (t) \ +\ q^2(1-t^2)(1+q^4t^2)(1+q^2+q^2t^2)\cdot L_{[21]} (t)
\ +\nn\\
\ + \  q^6(1-t^4)(1-t^6)\Big(1+q^2t^2+q^4t^2(1+t^2)+q^6t^6 \Big) \cdot L_{[111]} (t) \nn \\ \nn\\
(t/A)^{16}\cdot P^{[3,8]}_{_\Box}\ = \ L_{[3]} (t) \ +\
q^2(1-t^2)\Big(1+q^2(1+t^2)+q^4t^2+q^6t^2(1+t^2)+q^8t^4\Big)\cdot L_{[21]} (t)\ +\nn\\
\ + \  q^6(1-t^4)(1-t^6)\Big(1+q^2t^2+q^4t^2(1+t^2)+q^6t^4(1+t^2)+q^8t^8\Big)\cdot L_{[111]} (t)\nn\\
\ldots
\ee
Note that in all these examples, the power of polynomial in $q$ (but not in $t$!)
is exactly equal to that
of the Alexander polynomial, i.e. is regulated by the {\it defect} \cite{def}
of the differential expansion \cite{diffexpan}.
If true beyond these examples, this equality can provide a clue to Hall-Littlewood
expansions for non-torus knots.
It also deserves mentioning, that above expansions also hold beyond the topological locus,
i.e. for {\it extended} knot polynomials of \cite{DMMSS,MMMkn1}.

\bigskip

Taken literally, the suggestion in this letter implies that $p$-adic knot polynomials
are deducible from the ordinary superpolynomials, at least for torus knots in the
fundamental representation.
A really interesting question would be to extend this definition from torus to
generic knots or, to begin with, to some other simple families, like
twist, rational (two-bridge), pretzel, arborescent (double-fat) and fingered 3-strand
knots (see \cite{knotebook} for definitions and references).
In other words, one should look for an independent $p$-adic definition of
something like the $q$-expansion coefficients in superpolynomials,
which will be topological invariant (or Reidemeister invariant, if knots are
substituted by closed braids).
This should be done despite the failure of previous attempt to find $p$-adic
counterparts of ${\cal R}$-matrices.
Perhaps, the situation can be similar to that with the virtual knots \cite{virt},
where ${\cal R}$-matrix formalism {\it a la} \cite{RT,RTmod} also breaks down --
but just a little, so that topological invariant HOMFLY polynomials can still
be successfully defined \cite{virtHOMFLY}.
In the worst case these studies will further clarify the possible role of
Hall-Littlewood expansions of knot polynomials, which remain under-investigated
even in the case of torus knots.

\bigskip

In fact, (\ref{RJ2332}) can be alternatively rewritten \cite{HL} as
\be
\frac{A^3}{t^3}L_{[2]}(t^3)^* = \frac{A^4}{t^4}L_{[3]}(t^4)^*
\ee
where one first substitutes the power of $t$ instead of $t$ into (\ref{Lpols})
and then substitute (\ref{tolo}) for $p_k$, without changing $t$ --
this is not the same as simply changing $t$ in (\ref{Ldims}),
therefore we put the label $*$ into a different place.
This is a very different representation from (\ref{23HL}) and its $p$-adic
interpretation is somewhat less straightforward -- still it also deserves attention
in the search for $p$-adic knot invariants.
Again, the equality
\be
(t/A)^{(m-1)n}\cdot H^{[m,n]}_{_\Box}\{t|p_k\} = L_{[m]}\{t^n|p_k\}
\ee
holds for {\it extended} HOMFLY polynomials and not only on the topological locus.

\bigskip

To summarize, building $p$-adic HOMFLY polynomials seems to be within reach,
but it is a certain challenge for $p$-adic string theory,
and work in this direction would clearly be important for a variety of
branches of modern mathematical physics.

\section*{Acknowledgements}

I am indebted to A.Mironov for stimulating conversations on $p$-adic knot invariants
and to the organizers of Dragovich fest for a stimulus to write this letter.

This work was performed at the Institute for Information Transmission Problems with the financial
support of the Russian Science Foundation (Grant No.14-50-00150).


\begin{thebibliography}{12}


\bibitem{HL}
A.Mironov, A.Morozov, S.Shakirov and A.Sleptsov,
 JHEP 2012 (2012) 70,  arXiv:1201.3339;
 J. Phys. A: Math. Theor. {\bf 45} (2012) 355202,  arXiv:1203.0667  \\
P.Etingof, E.Gorsky and I.Losev,
arXiv:1304.3412 \\
Sh.Shakirov,  arXiv:1308.3838





\bibitem{pstr}
I.Volovich,
Class. Quant. Grav. {\bf 4} (1987) L83-L87 \\
Yu.I. Manin
{\it Reflections on arithmetical physics},
Poiana Brasov Proceedings, {\it Conformal invariance and string theory} (1987) 293-303;
P-adic Numbers {\bf 5} 313-325,  arXiv:1312.5160 \\
I.Aref'eva, B.Dragovic and I.Volovich,
Phys. Lett. {\bf B 200} (1988) 512-514;
Phys.Lett. {\bf B 209} (1988) 445-450;
Phys.Lett. {\bf B 212} (1988) 283-291;
Phys.Lett. {\bf B 214} (1988) 339-349 \\
A.Zabrodin,  Comm.Math.Phys. {\bf 123} (1989) 463 \\
A.Levin and A.Morozov, Phys.Lett. B243 (1990) 207-214\\
B.Dragovic,
Phys.Lett. {\bf B 256} (1991) 392-39;
J.Math.Phys. {\bf 34} (1992) 1143-1148,  arXiv:math-ph/0402037;
Theor.Math.Phys. {\bf 93} (1993) 1225-1231;
Theor.Math.Phys. {\bf 100} (1994) 1055-1064 \\
L.Brekke and P.Freund,
Phys. Rep. {\bf 233} (1993) 1-66 \\
V.Vladimirov, I.Volovich and E.Zelenov,
{\it p-Adic Analysis and Mathematical Physics}, WS, 1994 \\
G.Djordjevic, B.Dragovich and Lj.Nesic,
Infin.Dim.Anal.Quant.Prob.Relat.Top. {\bf 6} (2003) 179-195, hep-th/0105030;
Mod.Phys.Lett. {\bf A14} (1999) 317-325, (1999) hep-th/0005216 \\
B. Dragovich, A.Khrennikov, S.Kozyrev and I.Volovich, Funk.Anal.Appl. 1 (2009) 1-17,
arXiv:0904.4205



\bibitem{adelVen}
L.Brekke, P.Freund, M.Olson and E.Witten,  Nucl. Phys. {\bf B302} (1988) 365

\bibitem{LeMo}
L.Chekhov, A.Mironov and A.Zabrodin,
Commun.Math.Phys. {\bf 125} (1989) 675 \\
D.Lebedev and A.Morozov,
Theor.Math.Phys. 82 (1990) 1-6

\bibitem{knotpols}
J.W.Alexander, 
Trans.Amer.Math.Soc. {\bf 30} (2) (1928) 275-306;\\
J.H.Conway, 
Algebraic Properties,
In: John Leech (ed.), {\sl Computational Problems in Abstract Algebra}, Proc.
Conf.
Oxford, 1967, Pergamon Press, Oxford-New York, 329-358, 1970;\\
V.F.R.Jones, 
Invent.Math. {\bf 72} (1983) 1
Bull.AMS {\bf 12} (1985) 103
Ann.Math. {\bf 126} (1987) 335;\\
L.Kauffman,
Topology {\bf 26} (1987) 395;\\
P.Freyd, D.Yetter, J.Hoste, W.B.R.Lickorish, K.Millet,
A.Ocneanu,
Bull. AMS. {\bf 12} (1985) 239;\\
J.H.Przytycki and K.P.Traczyk,
Kobe J. Math. {\bf 4} (1987) 115-139

\bibitem{RTmod}
A.Mironov, A.Morozov and And.Morozov, JHEP 03 (2012) 034, arXiv:1112.2654 \\
S.Nawata, P.Ramadevi, Zodinmawia, J.Knot Theory and Its Ramifications 22 (2013) 13, arXiv:1302.5144 \\
A.Anokhina and And.Morozov, Teor.Mat.Fiz. 178 (2014) 3-68, arXiv:1307.2216 \\
A. Mironov and A. Morozov,  Nucl.Phys. B899 (2015) 395-413, arXiv:1506.00339 \\
A.Mironov, A.Morozov, An.Morozov and A.Sleptsov,  arXiv:1508.02870


\bibitem{knotebook}
http://knotebook.org

\bibitem{CS}
S.Chern and J.Simons, Proc.Nat.Acad.Sci. 68 (1971) 791794; Annals of Math. 99 (1974) 48-69 \\
A.S.Schwarz, New topological invariants arising in the theory of quantized fields, Baku Topol. Conf., 1987\\
M.Atiyah, The geometry and physics of knots, (CUP, 1990)



\bibitem{Wit} E.Witten, Comm.Math.Phys. {\bf 121} (1989) 351

\bibitem{MSm} A.Morozov and A.Smirnov, Nucl.Phys. {\bf B835} (2010) 284-313, arXiv:1001.2003 \\
A.Smirnov, Proc. of International School of Subnuclar Phys. Erice, Italy, 2009, arXiv:hep-th/0910.5011

\bibitem{CFT}  A.Belavin, A.Polyakov, A.Zamolodchikov, Nucl.Phys. B241 (1984) 333-380\\
A.Zamolodchikov and Al.Zamolodchikov,
{\it Conformal field theory and critical phenomena in 2d systems}, 2009 \\
A.Mironov, S.Mironov, A.Morozov and An.Morozov,
 Theor.Math.Phys. 165 (2010) 1662-1698,
 arXiv:0908.2064

\bibitem{inds}
R.K.Kaul, T.R.Govindarajan, Nucl.Phys. B380 (1992) 293-336, hep-th/9111063; B393 (1993) 392-412 \\
P.Ramadevi, T.R.Govindarajan and R.K.Kaul, Nucl.Phys. B402 (1993) 548-566, hep-th/9212110;
Nucl.Phys. B422 (1994) 291-306, hep-th/9312215; Mod.Phys.Lett. A10 (1995) 1635-1658, hep-th/9412084 \\
P.Ramadevi and Zodinmawia, arXiv:1107.3918; arXiv:1209.1346 \\
D. Galakhov,   D. Melnikov, A. Mironov, A. Morozov and A. Sleptsov,
Phys.Lett. B743 (2015) 71-74, arXiv:1412.2616;
JHEP 1507 (2015) 069,  arXiv:1412.8432;
Nucl.Phys. B899 (2015) 194-228,  arXiv:1502.02621\\
A.Mironov,  A.Morozov,   And.Morozov,   P.Ramadevi, V.K.Singh,
JHEP 1507 (2015) 109,  arXiv:1504.00371


\bibitem{WZNW}
E.Witten, Comm.Math.Phys. {\bf 92} (1984) 455 \\
V.Knizhnik and A.Zamolodchikov, Nucl.Phys. {\bf B247} (1984) 83

\bibitem{GMMOS}
M.Wakimoto, Commun.Math.Phys. 104 (1986) 605-609 \\
A.Morozov, JETP Lett. 49 (1989) 345-349 \\
A.Gerasimov, A.Marshakov, A.Morozov, M.Olshanetsky and S.Shatashvili,
Int.J.Mod.Phys. A5 (1990) 2495-2589

\bibitem{FZ}
P.Freund and A.Zabrodin,
Comm.Math.Phys. {\bf 147} (1992) 277-294, hep-th/9110066;
Phys.Lett. {\bf B294} (1992) 347-353, hep-th/9208063

\bibitem{IMMMfe}
H.Itoyama, A.Mironov, A.Morozov and An.Morozov, IJMP A28 (2013) 1340009, arXiv:1209.6304 \\
H.Fuji, S.Gukov and  P.Sulkovski,  arXiv:1205.1515

\bibitem{Gar}
S.Garoufalidis, S.Garoufalidis and T.Le,
 Geometry and Topology, {\bf 9}  (2005) 1253-1293, math/0309214; math/0503641 \\
A.Mironov and A.Morozov,  AIP Conf.Proc. {\bf 1483} (2012) 189-211,  arXiv:1208.2282 \\
S.Garoufalidis, P.Kucharski and P.Sulkowski,  arXiv:1504.06327

\bibitem{AMMEO}
 A.Alexandrov, A.Mironov and A.Morozov, Int.J.Mod.Phys. A19 (2004) 4127-4165, arXiv:hep-th/0310113;
Int.J.Mod.Phys. A21 (2006) 2481-2518, hep-th/0412099;
Fortsch.Phys. 53 (2005) 512-521, arXiv:hep-th/0412205;
Teor.Mat.Fiz. 150 (2007) 179-192, hep-th/0605171;
Physica D235 (2007) 35 126-167, hep-th/0608228;
JHEP 0912 (2009) 053, arXiv:0906.3305 \\
L.Chekhov, B.Eynard and N.Orantin, JHEP12 (2006) 053, math-ph/0603003 \\
B.Eynard and N.Orantin, Commun. Number Theory Phys., 1 (2007) 347-452, math-ph/0702045 \\
R.Dijkgraaf, H.Fuji and M.Manabe, Nucl.Phys. {\bf B849} (2011) 166-211,  arXiv:1010.4542 \\
J.Gu, H.Jockers, A.Klemm and  M.Soroush,  arXiv:1401.5095


\bibitem{DMMSS}
P.Dunin-Barkowski, A.Mironov, A.Morozov, A.Sleptsov, A.Smirnov, JHEP 03 (2013) 021, arXiv:1106.4305

\bibitem{RJ}
M.Rosso and V.F.R.Jones, J. Knot Theory Ramifications, 2 (1993) 97-112 \\
X.-S.Lin and H.Zheng, Trans. Amer. Math. Soc. 362 (2010) 1-18 math/0601267


\bibitem{MD}
 I.Macdonald, {\it Symmetric Functions and Hall Polynomials}, 2nd ed. Oxford University Press,
New York, 1999

\bibitem{suppo}
M.Khovanov. Duke Math.J. 101 (2000) no.3, 359426, math/9908171 \\
S.Gukov, A.Schwarz and C.Vafa, Lett.Math.Phys. 74 (2005) 53-74, hep-th/0412243 \\
M.Khovanov and L.Rozansky, Fund. Math. 199 (2008), no. 1, 191, math/0401268; Geom.Topol. 12 (2008),
no. 3, 13871425, math/0505056; math/0701333 \\
V.Dolotin and A.Morozov,  Nucl.Phys. {\bf B878} (2014) 12-81,  arXiv:1308.5759

\bibitem{phs}
F.Mautner, Am.J.Math. {\bf 80} (1958) 441\\
P.Cartier, Proc.Symp.Pure Math. {\bf 26} (1973) AMS Providence \\
P.Freund,  Phys. Lett. {\bf 257B} (1991) 119


\bibitem{def} Ya.Kononov and A.Morozov, JETP Lett. {\bf 101} (2015) 12, 831-834,
 arXiv:1504.07146

\bibitem{diffexpan}
N.Dunfield, S.Gukov and J.Rasmussen, Experimental Math. 15 (2006) 129-159, math/0505662 \\
A.Mironov, A.Morozov and And.Morozov, AIP Conf.Proc. 1562 (2013) 123-155, arXiv:1306.3197 \\
S.Arthamonov, A.Mironov, A.Morozov, And.Morozov,
Theor.Math.Phys. 179 (2014) 509, arXiv:1306.5682

\bibitem{MMMkn1}
A.Mironov, A.Morozov and And.Morozov, in: {\it Strings, Gauge Fields, and the Geometry Behind: The
Legacy of Maximilian Kreuzer}, edited by A.Rebhan, L.Katzarkov, J.Knapp, R.Rashkov, E.Scheidegger
(World Scietific Publishins Co.Pte.Ltd. 2013) pp.101-118, arXiv:1112.5754

\bibitem{RT}
N.Yu.Reshetikhin and V.G.Turaev, Comm. Math. Phys. 127 (1990) 1-26 \\
E.Guadagnini, M.Martellini and M.Mintchev, Clausthal 1989, Procs. 307-317; Phys.Lett. B235 (1990) 275

\bibitem{virt}  L.H.Kauffman, Eur.J.Comb. 20 (1999) 663-690, math/9811028 \\
R.Fenn, D.P.Ilyutko, L.H.Kauffman and V.O.Manturov, arXiv:1409.2823

\bibitem{virtHOMFLY}
L.Bishler, A.Morozov, An.Morozov and Ant.Morozov,  Phys.Lett. B737 (2014) 48-56, arXiv:1407.6319;
 Int.J.Mod.Phys. A30 (2015) 1550074, arXiv:1411.2569 \\
A.Morozov, An.Morozov and A.Popolitov,
Phys.Lett. B749 (2015) 309-325, arXiv:1506.07516; arXiv:1508.01957

\end{thebibliography}
\end{document}